\documentclass[12pt,a4paper]{article}

\usepackage{times} 
\usepackage[a4paper, lmargin=2cm, rmargin=2cm , tmargin=2.5cm , bmargin=2.25cm  ,footskip=1.25cm]{geometry} 
\usepackage{graphicx} 
\usepackage[square,sort,comma,numbers]{natbib} 
\usepackage{amsmath} 
\usepackage[hyphens]{url} 
\usepackage{soul}
\usepackage{fancyhdr} 
\usepackage[explicit]{titlesec} 
\usepackage{abstract} 
\usepackage{authblk} 
\usepackage{tabularx}
\usepackage{multirow} 
\usepackage{colortbl} 
\usepackage{booktabs}
\usepackage{array}
\usepackage[super]{nth} 
\usepackage{algorithm}
\usepackage[end]{algpseudocode}

\usepackage{subfig}

\usepackage{fp}
\usepackage{tikz}
\usepackage{xcolor}
\usetikzlibrary{arrows}
\usetikzlibrary{decorations.pathmorphing}
\usetikzlibrary{decorations.pathreplacing}
\usetikzlibrary{calc,patterns,decorations.markings}
		\definecolor{darkblue}{rgb}{0.0, 0.0, 0.55} 
\usetikzlibrary{shapes,positioning,calc}

\graphicspath{{figures/}}

\titleformat{\section}{\normalfont\bfseries}{\thesection}{1em}{\MakeUppercase{#1}}  

\titleformat{\subsection}{\normalfont\bfseries\small}{\thesubsection}{1em}{{#1}} 

\titleformat{\subsubsection}{\normalfont\small}{\thesubsubsection}{1em}{{#1}} 

\algrenewcommand\algorithmicend{\textbf{end}}
\algrenewtext{EndFor}{\algorithmicend}

\makeatletter
\newcommand{\thickhline}{ \noalign {\ifnum 0=`} \fi \hrule height 1pt \futurelet \reserved@a \@xhline }
\newcommand{\morethickhline}{ \noalign {\ifnum 0=`} \fi \hrule height 2pt \futurelet \reserved@a \@xhline }
\newcolumntype{"}{@{\hskip\tabcolsep\vrule width 1pt\hskip\tabcolsep}}
\makeatother

\newlength{\Oldarrayrulewidth}

\begin{document}

\title{\vspace{-1.3cm}\normalsize\normalfont\bfseries \MakeUppercase { Model Based Active Slosh Damping Experiment} }
\date{}
\author[1]{C. Jetzschmann}
\author[1]{H. Strauch}
\author[2]{S. Bennani}
\affil[1]{Airbus DS GmbH, Airbus Allee 1, 28199 Bremen, Germany

christina.jetzschmann@airbus.com +494215393597, hans.strauch@airbus.com +494215395316}
\affil[2]{ESA/ESTEC, Keplerlaan 1, 2200 AG Noordwijk, The Netherlands 

samir.bennani@esa.int +31715658099}

\renewcommand\Authands{, }
\renewcommand{\Authfont}{\normalsize\normalfont \bfseries}
\renewcommand{\Affilfont}{\normalsize\normalfont}
\renewcommand{\abstractnamefont}{\bfseries\normalsize\MakeUppercase} 

\maketitle

\begin{abstract}
This paper presents a model based experimental investigation to demonstrate the usefulness of an active damping strategy to manage fluid sloshing motion in spacecraft tanks. The active damping strategy is designed to reduce the degrading impact on maneuvering and pointing performance via a force feedback strategy. Many problems have been encountered until now, such as instability of the closed loop system, excessive consumption in the attitude propellant or problems for engine re-ignition in upper stages. Mostly, they have been addressed in a passive way via the design of baffles and membranes, which on their own have mass and constructive impacts. Active management of propellant motion in launchers and satellites has the potential to increase performance on various levels. 
This paper demonstrates active slosh management using force feedback for the compensation of the slosh resonances. Force sensors between tank and the carrying structure provide information of the fluid motion via the reaction force. The control system is designed to generate an appropriate acceleration profile that leads to desired attenuation profiles in amplitude, frequency and time.
Two robust control design methods, one based on $\mu$ design and the other on parametric structured design based on non-smooth optimization of the worst-case $H_{\infty}$ norm, are applied.
The controller is first tested with a computational fluid dynamics simulation in the loop. Finally a water tank mounted on a Hexapod with up to $1100$ liter is used to evaluate the control performance.
The paper illustrates that is possible to actively influence sloshing via closed loop.
\end{abstract}

\thispagestyle{fancy}

\section{Introduction}
The ratio between propellant mass and dry mass in satellites and particularly in launchers can easily exceed one.
Such a high value of fluid can cause instability of the closed loop system, excessive consumption in the attitude propellant when the control system tries to counteract the sloshing, or problems for engine re-ignition in upper stages because the propellant is not properly located at the tank outlet. 
In most of the cases this has been only addressed in a passive fashion: On the constructive side  baffles or membranes
are implemented. On the flight software side (GNC level) the destabilizing impact of the sloshing resonant mode is often just filtered out via notch filters. This is considered as a passive approach since there is no sensing, actuation and control action.

Active management of propellant motion offers the potential to increase performance on various levels. Examples are large angle slewing maneuvers of upper stages performed in such a way that the propellant stays close to the tank bottom or increasing the damping in powered flight phases of launchers. Force sensors between tank and the carrying structure can provide information of the fluid motion via the reaction force. A control system will be designed generating an appropriate acceleration profile that leads to a desired closed loop behavior. 

\begin{figure}[!htb]
\begin{center}
\includegraphics[width=0.8\columnwidth]{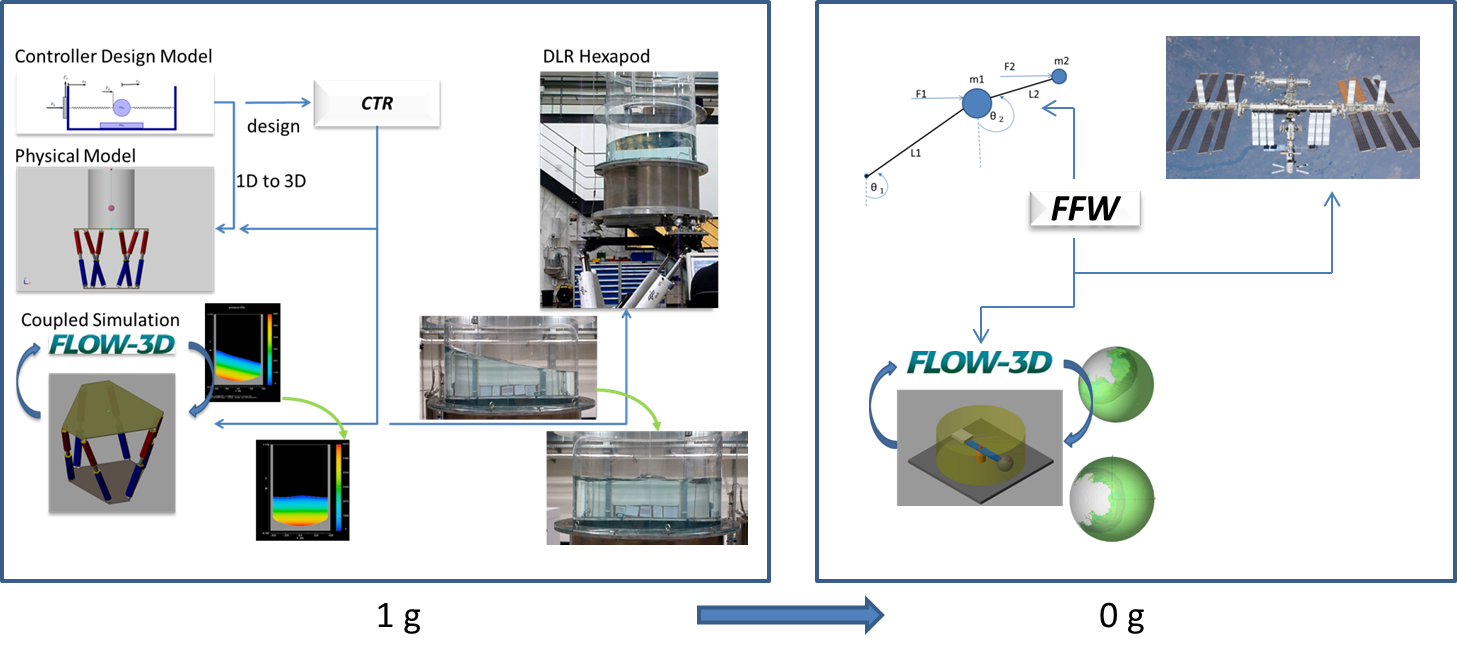}
\caption{Workflow Development and Test - Left side: Active damping on Hexapod under one g, Right side: Excitation free spin-up (open loop) experiment on ISS in zero g (in preparation)}
\label{fig:overview_1}
\end{center}
\end{figure} 

A model based design approach allows approaching the development of such a controller in a structured way (\cite{Strauch2014}). At first, an analytic description of the slosh motion is derived. This serves as the basis for the control design. The next step is to set-up a simulation infrastructure based on physical modeling, in this case with Flow3D for a computational fluid dynamic (CFD) representation of the propellant dynamics (\cite{Fries2012}) and Mathwork's SimMechanics for the Hexapod representation which is used for
exciting the slosh motion. It is of importance that the CFD must interact with the controller in closed loop (\cite{Behruzi2016}). Replacing the analytic design model with the CFD provides a first test on the suitability of the designed controller. The last step must of course be the test with real fluids.

In order to prove the concept of active damping and to show the mastering of the complete development cycle from theoretical design to CFD simulation to hardware-in-the-loop, a tank with fill levels from $600$ to $1100$ liter water is placed on a Hexapod. The fluid is deliberately excited into an oscillatory phase. The reaction force between tank and Hexapod attachment is measured and the controller commands the Hexapod with the purpose of damping the slosh motion after it has been excited into an oscillatory, very lightly damped motion.

Two controller architectures are tested and compared. The first is a classical $\mu$ design with a model reduction applied to the resulting controller. The second one is a fixed structure controller based on the Mathworks' Robust Control Toolbox 'robust systune' feature, which implements the algorithm from \cite{Apkarian2015}. This last technique is very attractive insofar as it allows to derive robust controller for prescribed controller topologies. However, the parametric structured design, which minimizes the worst-case $H_{\infty}$ norm, is based on non-smooth optimization. This can lead to large, abrupt changes in the controller parameters for even small changes in the weighting function, which defy the intuition. A full order $\mu$ controller is therefore a good guiding principal serving as a benchmark.

In order to have a reliable process allowing to move from simulation to the hardware-in-the-loop,
a process has been established which automatically generates a LabView implementation from the Simulink design environment allowing an effective way of implementing complex algorithms into embedded systems. 

The paper illustrates that is possible to actively influence sloshing via closed loop and that the algorithm described in
\cite{Apkarian2015} is very useful in designing robust feedback system if a prescribed controller structure is desired. Of course the limitation due to the $1$ g environment must be acknowledged. In fact, the described activity is a pre-cursor towards an ISS based experiment testing an excitation-free spin-up (see figure~\ref{fig:overview_1}).

The work has been performed under ESA's Future Launcher Preparatory Program (FLPP3) in the study {\it Upper Stage Attitude Control Development Framework} (USACDF). The support with the Hexapod tests by DLR Institute of Space Systems is gratefully acknowledged. More details of the Hexapod tests and the CFD simulations can be found in~\cite{Konopka2016}. 

\section{Controller Design Model}

\subsection{Differential Equation of Motion}

\begin{figure}[!ht]
\begin{center}
\begin{tikzpicture}[scale=0.8, transform shape]
\tikzstyle{spring}=[thick,decorate,decoration={zigzag,pre length=0.3cm,post length=0.3cm,segment length=6}]
\node (forceh) 	[yshift=2.0cm,xshift=-0.55cm] {};	
\node (forces) 	[yshift=2.85cm,xshift=-0.25cm] {};	
\node (s1)		[yshift=2.0cm,xshift=-0.05cm] {};	
\node (s2)		[yshift=2.0cm,xshift=10.1cm] {};	
\node (center)  [yshift=2.0cm,xshift=5cm] {};		
\node (centerl) [yshift=2.0cm,xshift=4.2cm] {};		
\node (centerr) [yshift=2.0cm,xshift=5.8cm] {};		
\node (Xh)		[yshift=4.15cm,xshift=-0.15cm] {};	
\node (Xs)		[yshift=4.15cm,xshift=5cm] {};		
\node (Fd)		[yshift=3.5cm,xshift=4.2cm] {};		
\node (M1) [minimum width=0.5cm,minimum height=2cm,yshift=2cm,xshift=-0.2cm,style={draw=darkblue,fill=black!30!white,outer sep=0pt,thick=1mm}] {};
\draw [line width=1.5pt, ->, >=latex](-2,2) -- (forceh.east) node [anchor=south,xshift=-1cm] {$F_h$};
\draw [line width=1.5pt, ->, >=latex](forces.north) -- (-0.25,4.5) node [anchor=south,xshift=0cm] {$F_s$};
\draw [spring](s1) -- (centerl.center) node [pos=0.5,above] {};
\draw [spring](centerr.center) -- (s2) node [pos=0.5,above] {};
\draw [style={draw=darkblue,fill=blue!30!white,outer sep=0pt,thick=1mm}](center) circle(0.8cm) node {$m_s$};
\draw[line width=2mm,color=darkblue] (0,4 ) -- (0,0) -- (10,0) -- (10,4);
\draw [line width=1.5pt, |->, >=latex](Xh.east) -- (1.7, 4.15) node [anchor=south,xshift=-0.2cm] {$x_h$};
\draw [line width=1.5pt, |->, >=latex](Xs.center) -- (6.7, 4.2) node [anchor=south,xshift=-0.2cm] {$x_s$};
\draw [line width=0.8pt, - ,>=latex ](centerl.center) -- (4.2, 3.5);
\draw [line width=1.5pt, -> ,>=latex ](3.2, 3.5) -- (Fd.center) node [anchor=south,xshift=-0.5cm] {$F_d$};
\node (Mr) [minimum width=4cm,minimum height=0.5cm,yshift=0.35cm,xshift=5cm,style={draw=darkblue,fill=blue!30!white,outer sep=0pt,thick=1mm}] {$m_r$};
\end{tikzpicture}
\caption{Controller design model: Point mass respresenting the slosh mass $m_s$ attached in a movable frame with mass $m_r$ (comprising the mass of the frame as well as the non-sloshing mass of the fluid) with a spring/damper system. (The damping providing dash-pod is omitted in the picture)
The Hexapod system generates the external force $F_h$. The measurement used for control is a force sensor $F_s$ (negative sign on push, positive on pull).
A fictitious disturbance force $F_d$ is used for providing an input for the performance weightings needed in the $\mu$ design. 
}
\label{fig:design_model}
\end{center}
\end{figure}
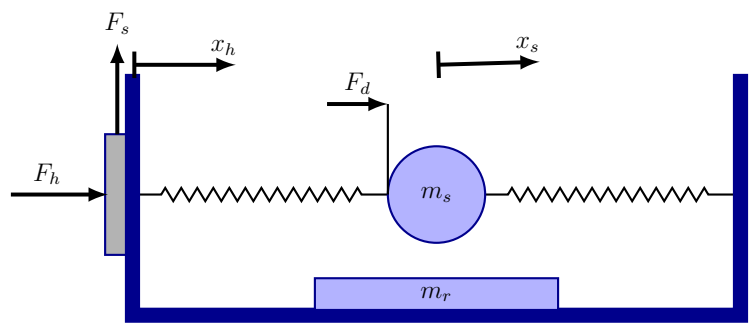

A CFD model cannot directly be used for a controller design. A dedicated design model is needed for any model based design technique. One goal of the activity was also to evaluate whether a simple model, as will be laid out next, can lead to a satisfactory design.
Figure~\ref{fig:design_model} shows the schematic for a 1D motion.

The control task is to damp the relative motion, expressed as 
$d/dt (x_s - x_h) \rightarrow 0$. It is assumed that a fictitious force $F_d$
can move the sloshing mass $m_s$. The goal of this force is to provide an artificial input which can be used to express performance requirements. The Hexapod position $x_h$ is in principle  accessible for a feedback, but the position $x_s$, for $m_s$ being purely
a fictitious object, not. The relative motion can therefore only be 
accessed indirectly via its reaction to the tank wall. This reaction force is measured by a force sensor $F_s$. Of course the same force sensor also captures the external force $F_h$ generated by the Hexapod drive as commanded by the controller. The feedback sensor therefore measures not only the slosh reaction force but also control command. This means that the D matrix is non-zero, yet control and sensor are at least collocated.
The fluid mass is broken down into two components. One moves in synchronization with the Hexapod itself and is therefore called $m_{rigid_{prop}}$. It essentially behaves like a rigid mass which is connected with the Hexapod . The other is sloshing part and named $m_s$. Based on figure~\ref{fig:design_model} and with

\begin{description}
\item[$m_r$] the mass of the container consisting of the tank and the moving part
of the Hexapod as well the "rigid" (non-sloshing) part of the fluid: $m_r = m_{hex} + m_{rigid_{prop}}$
\item[$m_s$] the sloshing mass
\item[$F_s$] the load sensor (negative sign on push, positive on pull)
\item[$F_h$] the force from the Hexapod drive
\item[$F_d$] an "auxiliary" force to be used in the design process for specifying the performance weights
\item[$k$, $c$] the spring constant and damping constant
\end{description}

the following holds:

\begin{equation}
m_s \ddot{x_s} = -k (x_s - x_h) - c \frac{d}{dt}(x_s - x_h) + F_d
\end{equation}

\begin{equation}
m_r \ddot{x_H} = k(x_s - x_h) + c \frac{d}{dt}(x_s - x_h) + F_h
\end{equation}

The sensor measures the rigid motion as well as the sloshing one:

\begin{equation}
F_s = m_r \ddot{x_H} -  k(x_s - x_h) - c \frac{d}{dt}(x_s - x_h)
\end{equation}\label{eq:Fs}

Let $x = [x_s \quad x_h  \quad \dot{x}_s \quad \dot{x}_h]^T$ be the state vector and $[F_d \quad F_h ]^T$ the input vector, then the $A, B, C, D$ matrices of the state space system are given as follows

\begin{align}
A = 
\begin{bmatrix}
	0 & 0 & 1 & 0\\
	0 & 0 & 0 & 1\\
	-k/m_s & k/m_s & -c/m_s & c/m_s\\
	k/m_r & -k/m_r & c/m_r & -c/m_r
\end{bmatrix}
\qquad
B = 
\begin{bmatrix}
	0 & 0\\
	0 & 0\\
	1/m_s & 0 \\
	0 & 1/m_r\\
\end{bmatrix}
\label{eq:A_mat}
\end{align}

With the output vector $y = [F_{sp} \quad \Delta \quad \dot{\Delta} \quad x_h \quad \dot{x}_h \quad \ddot{x}_h ]^T$, where $\Delta = (x_s - x_h)$ is the relative position of the slosh mass in the tank and $F_{sp}$ is the spring reaction force, the output matrix $C$ and $D$ are given as:

\begin{align}
C = 
\begin{bmatrix}
k & -k & c & -c\\
1 & -1 & 0 & 0\\
0 & 0 & 1 & -1\\
0 & 1 & 0 & 0\\
0 & 0 & 0 & 1\\
k/m_h & -k/m_h & c/m_h & -c/m_h
\end{bmatrix}
\qquad
D = 
\begin{bmatrix}
0 & 0\\
0 & 0\\
0 & 0\\
0 & 0\\
0 & 0\\
0 & 1
\end{bmatrix}
\label{eq:C}
\end{align}

Of course  $\Delta$ and $\dot{\Delta}$ is not actually observable, but the output is needed in the $\mu$ design for performance weighting. The sensor output in terms of the output vector $y$ of the differential equation is given as the sum of the spring reaction force $F_{sp}$ and the rigid body force $m_h \ddot{x}_h$:

\begin{equation}
F_s = F_{sp} +  m_h \cdot \ddot{x_h}  = y(1) +  m_h  \cdot y(6)   
\end{equation}

However, the Hexapod drive system request a (consistent) set of position, velocity and acceleration commands for $x_h$. Therefore the controller command output must be the desired acceleration $\ddot{x}_{h_{cmd}}$. The velocity and position command to the Hexapod drive are then computed from  $\ddot{x}_{h_{cmd}}$ by integrating twice. A lowpass filter with uncertain time delay is used to cover the drive loop. This will be discussed in the following structured uncertainty modeling subsection.
\subsection{Structured Uncertainty Modelling}
\begin{figure}[!htb]
\begin{center}
\begin{tikzpicture}[scale=0.8, transform shape, >=stealth] 
\tikzstyle{sum} = [draw, circle, node distance=1.5cm]
  \coordinate (orig)   	at (0,0);
  
  \coordinate (A0)	   	at (1,4);
  \coordinate (Aa) 		at (2.77,4.75);
  \coordinate (Ab) 		at (3.0,4.75);
  \coordinate (Ac) 		at (0.75,4.75);
  \coordinate (Ad) 		at (1,4.75);
  \coordinate (Ae) 		at (0.25,4.5);
  \coordinate (Af) 		at (0.25,4.25);
  \coordinate (Ag) 		at (1,4.25);
  
  \coordinate (C0) 		at (3.5,4);
  \coordinate (Ca) 		at (5.47,4.75);
  \coordinate (Cb) 		at (5.6,4.75);
  \coordinate (Cc) 		at (3.25,4.75);
  \coordinate (Cd) 		at (3.5,4.75);
  \coordinate (Ce) 		at (5.47,4.55);
  \coordinate (Cf) 		at (5.47,4.35);
  \coordinate (Cg) 		at (5.47,4.15);
  \coordinate (Ch) 		at (6.3,4.55);
  \coordinate (Ci) 		at (5.6,4.15);
  \coordinate (Cj) 		at (14.6,2.0);
  
  \coordinate (E0) 		at (7.0,4.5);
  \coordinate (F0) 		at (7.0,3.4);
  \coordinate (Fa) 		at (7.0,4.0);
  \coordinate (Fb) 		at (6.8,3.6);
  \coordinate (Fc) 		at (7.0,3.6);
  \coordinate (Fd) 		at (8.75,3.6);
  \coordinate (Fe) 		at (8.9,3.6);
  
  \coordinate (G0) 		at (7.5,2.5);
  
  \coordinate (H0) 		at (7.0,5.9);
  \coordinate (Ha) 		at (6.6,5.7);
  \coordinate (Hb) 		at (9.8,4.35);
  \coordinate (Hc) 		at (7.0,6.2);
  \coordinate (Hd) 		at (5.5,7.1);
  \coordinate (He) 		at (7.0,7.1);
  
  \coordinate (Ia) 		at (8.77,6.9);
  \coordinate (Ib) 		at (8.5,6.9);
  \coordinate (Ic) 		at (10.53,6.9);
  \coordinate (Id) 		at (10.77,6.9);
  
  \coordinate (Ja) 		at (11.55,6.9);
  \coordinate (Jb) 		at (11.3,6.9);
  \coordinate (Jc) 		at (13.3,6.9);
  \coordinate (Jd) 		at (13.5,6.9);
  \coordinate (Je) 		at (14.6,5.7);
  
  \coordinate (d1)	   	at (4.0,5.5);
  \coordinate (d2)	   	at (1.5,5.5);
  \coordinate (d4)	   	at (9.2,7.5);
  \coordinate (d5)	   	at (12.0,7.5);

  \coordinate (S1) 		at (6.6,4.35);
  \coordinate (S1a) 	at (6.6,4.89);
  \coordinate (S1b) 	at (6.6,4.0);
  \coordinate (S2) 		at (9.3,4.35);
  \coordinate (S3) 		at (11.3,6.65);

  \node[draw, minimum width=1.5cm, minimum height=1.0cm, anchor=south west, text width=1.5cm, align=center] (A) at (A0) {$LP_1$};
  \node[draw, minimum width=0.5cm, minimum height=0.5cm, anchor=south west, text width=0.5cm, align=center] (B) at (d2) {$\Delta_2$};
  \node[draw, minimum width=1.7cm, minimum height=1.0cm, anchor=south west, text width=1.7cm, align=center] (C) at (C0) {$DEQ$};
  \node[draw, minimum width=0.5cm, minimum height=0.5cm, anchor=south west, text width=0.5cm, align=center] (D) at (d1) {$\Delta_1$};
  \node[draw, minimum width=1.5cm, minimum height=0.75cm, anchor=south west, text width=1.5cm, align=center] (E) at (E0) {$m_s$};	   
  \node[draw, minimum width=1.5cm, minimum height=0.75cm, anchor=south west, text width=1.5cm, align=center] (F) at (F0) {$m_p$};
  \node[draw, minimum width=0.5cm, minimum height=0.5cm, anchor=south west, text width=0.5cm, align=center] (G) at (G0) {$\Delta_3$};
  \node[sum] (sum1) at (S2) {\tiny{+}};
  \node[draw, minimum width=0.5cm, minimum height=1.5cm, anchor=south west, text width=0.5cm, align=left] (H) at (H0) {\tiny{+\\[0.2cm]+\\+}};
  \node[text=red](out1) at (Cj){\hspace{0.5cm}$p$};  
  \node[draw, minimum width=1.5cm, minimum height=1.0cm, anchor=south west, text width=1.5cm, align=center,right=of H] (I){$LP_2$};
  \node[draw, minimum width=1.5cm, minimum height=1.0cm, anchor=south west, text width=1.5cm, align=center,right=of I] (J){$e^{-ST}$};
  \node[draw, minimum width=0.5cm, minimum height=0.5cm, anchor=south west, text width=0.5cm, align=center] (K) at (d4) {$\Delta_4$};
  \node[draw, minimum width=0.5cm, minimum height=0.5cm, anchor=south west, text width=0.5cm, align=center] (L) at (d5) {$\Delta_5$};
  \node[text=red](out2) at (Je){$c$};  
  \node[text=blue,right=of J](out3){$F_h$}; 
  
  \draw[->] (A.east) -- (C.west);
  \draw[-]  (Aa) |- (Ab);
  \draw[->] (Ab) |- (B.east);
  \draw[-]  (B.west) -| (Ac);
  \draw[->] (Ac) -- (Ad);
  \draw[red,->] (Ae) -- node[above left]{$F_d$}(A.west);
  \draw[blue,->] (Af) -- node[below left]{$\ddot{x}_{cmd}$}(Ag);
  
  \draw[-]  (Ca) |- (Cb);
  \draw[->] (Cb) |- (D.east);
  \draw[-]  (D.west) -| (Cc);
  \draw[->] (Cc) -- (Cd);
  \draw[red,-]  (Cg) |- (Ci);
  \draw[red,->] (Ci) |- (Cj);

  \draw[-]  (Cf) -- node[below right]{$\ddot{x}_{h}$}(S1);
  \draw[-]  (S1) -- (S1a);
  \draw[->] (S1a) -- (E.west);
  \draw[-]  (S1) -- (S1b);
  \draw[->] (S1b) -- (Fa); 

  \draw[-]  (G.west) -| (Fb);
  \draw[->] (Fb) -- (Fc);
  \draw[-]  (Fd) -- (Fe);
  \draw[->] (Fe) |- (G.east);
  
  \draw[->] (E) -| (sum1);
  \draw[->] (F) -| (sum1);
  
  \draw[-]  (Ce) -- node[above]{\hspace{2mm}$F_h$}(Ch);
  \draw[->] (Ch) |- (H.west);
  
  \draw[-]  (sum1) -- (Hb);
  \draw[-]  (Hb) |- (Ha);
  \draw[->] (Ha) |- (Hc);
  \draw[red,->] (Hd) -- node[above left]{$n$}(He);
  
  \draw[->] (H) -- (I);
  \draw[->] (I) -- (J);
  \draw[->] (Ib) -- (Ia);
  \draw[-]  (Ib) |- (K.west);
  \draw[-]  (Ic) -- (Id);
  \draw[->] (Id) |- (K.east);
  
  \draw[->] (Jb) -- (Ja);
  \draw[-]  (Jb) |- (L.west);
  \draw[-]  (Jc) -- (Jd);
  \draw[->] (Jd) |- (L.east);
  
  \draw[red,->]  (S3) |- (out2);
  \draw[blue,->]  (J.east) -- (out3);
  
\end{tikzpicture}

\caption{LFT model with input/output for performance weighting (red). Input: $n$ (measurement noise), $F_d$ (disturbance). Outputs: $p$ (relative velocity), $c$ (control effort). The controller input/output (blue) are $\ddot{x}_{cmd}$ (commanded Hexapod acceleration) and $F_h$ (force measurement). $LP_x$ - lowpass for hexapod drive and force sensor, $DEQ$ - differential equation (see equation~\ref{eq:A_mat}),
}
\label{fig:simul2}
\end{center}
\end{figure}
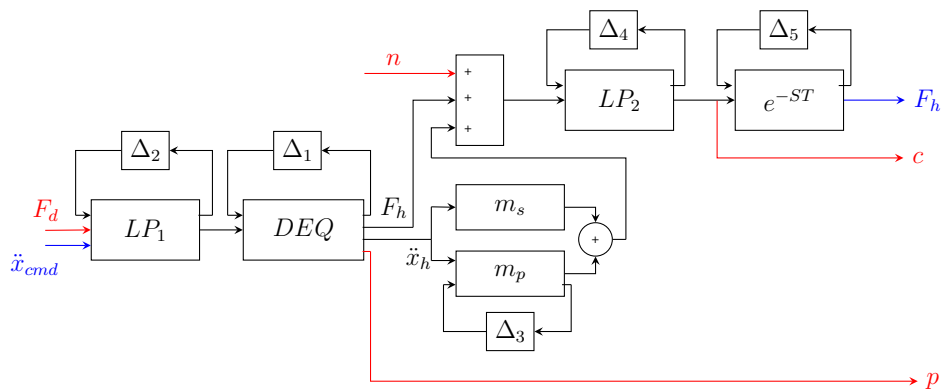

The following discusses the uncertain elements of the LFT (see figure~\ref{fig:simul2}).
Observing that $k$ and $m_s$ always comes as ration $k/m_s$, it should be equivalent to have just one uncertain element which is $k$ ($\Delta_1$ in figure~\ref{fig:simul2}) . The factor $c/m_s$ is not modeled as uncertainty. In fact the system is practically un-damped for the time frames relevant to the control task. Using therefore a very low (conservative) damping value for $c$ should be sufficient and no extra uncertainty modeling is needed. 
Another uncertain element is $m_r$ ($\Delta_3$). Obviously $m_r$ and $m_s$ must have a high uncertainty attached because these values are not directly accessible by physical measurements. ($m_s$ uncertainty was already lumped into $k$). 

The overall system certainly has time delays uncertainties. One time delay is placed at the actuator channel. In order to fit a delay into the LFT framework, a Pade approximation of second order is used. The corresponding state space system is implemented as an uncertain system. 

The actuator output channel as well as the input channel also gets a multiplicative, complex uncertainties ($\Delta_2$, $\Delta_4$ ), in order to cover un-modeled high dynamics and further making the $\mu$ framework mixed real/complex.
\subsection{System Identification} \label{sysid}
\begin{figure}[!htb]
\centering\fbox{
\includegraphics[width=0.7\textwidth]{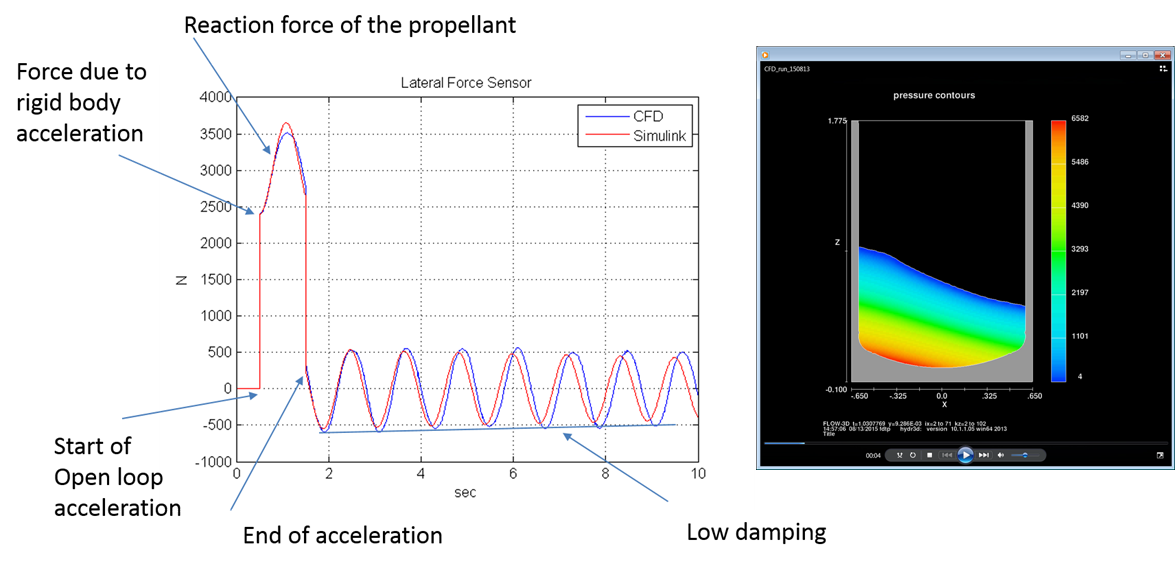}}
\caption{Comparison CFD response and Design model as reaction to a acceleration pulse.
}\label{fig:id1}
\end{figure}

In~\cite{dodge2000} estimates of $m_r$, $m_s$ and $k$ are given as a function of the tank form and the filling level. These values are cross checked by comparing a Flow3D simulation with the response of the design model. Figure~\ref{fig:id1} illustrates one such test used for adapting the model parameters of the design model. For $1.5 sec$ a constant acceleration is applied. The force computation from the CFD and the force sensor output from the design model immediately show the reaction force from the rigid part ($D$ matrix non-zero in the model) and then the force created by the moving fluid or by the spring of the design model. Once the acceleration stops only the fluid reaction is visible in the force.  

The immediate jump can be used to estimate $m_r$. The amplitude of the first peak gives information on $k$ and the frequency of the oscillation allows to estimate $k/m_s$. After fitting these three values figure~\ref{fig:id1} resulted. It can be seen that the rigid part $m_r = m_{hex} + m_{rigid_{prop}}$ is captured quite good. Actually, emphasis was put onto this part because the sensor will feedback to the controller its own command as force. Therefore an error in the rigid part force prediction will be attributed as slosh motion instead of getting attribute to the structure. 

The interesting part to be observed is that for the first three oscillations the frequency is well captured. Later the CFD frequency starts to become a little lower. It is obvious that the CFD motion can not completely be represented by a linear spring/damper model. However, the first three oscillations are captured quite well and this covers the time during which the controller shall damp the motion.

The parameters as taken from~\cite{dodge2000} were as initial guess quite accurate and  have only been modified by around $10 \%$. 
It is understood that above approach, based on only a couple of simulations for the particular fill level of the first test, is more curve fitting than system identification. This is justified by the interest to have at the first tank tests secure conditions by having a stable closed loop. As it turned out later this precaution was not necessary and the reliance on~\cite{dodge2000} for different fill level resulted in very good close loop performance for active damping. This statement holds not only for different fluid masses but also for different excitation level.

\section{Controller Design}
\subsection{Weighting Functions}
When approaching a controller design task the precision of the  specifications and the available knowledge can vary. At one end of the spectrum there may exist very good plant models including a good understanding of the associated uncertainties. At the same time the closed loop performance specification may also be  given in very precise quantitative terms. In such a case the LFT and the selection of the weighting functions are quite well determined by this knowledge and the specification.

\begin{figure}[!htb]
\centering\fbox{
\includegraphics[width=0.7\textwidth]{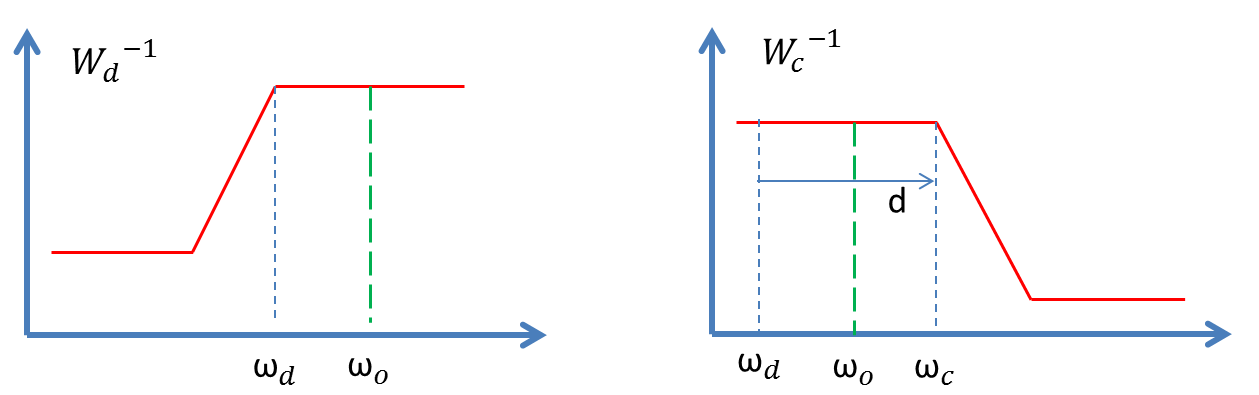}}
\caption{Inverse weighting functions for disturbance and control effort: $\omega_d$ - desired minimum frequency for disturbance rejection, 
$\omega_o$ - first sloshing mode, $\omega_c$ - desired roll-off frequency with relative distance $d$ from disturbance rejection frequency
}\label{fig:weightings}
\end{figure}

The active damping experiment does not belong into this category. One goal of the tests is exactly to find out how good the modeling matches with the actual behavior of the fluid when it comes down to closed loop design. Therefore the numerical values for the $\Delta$ blocks of figure~\ref{fig:simul2} are (at least for the first runs) rough guesses. A similar argument applies for the selection and the weightings of the performance inputs/outputs. Attention is put more on having robust stability (in the $\mu$ theory sense) than robust performance because at the beginning of the activity there did not exist a clear expectation of what can be achieved in terms of damping performance. As a consequence the selected input/outputs have a generic character. 

As can be seen by the red colored elements in figure~\ref{fig:simul2} one input is on the sensor model. This input shall cover the measurement noise. The other input is the disturbance force $F_d$ acting on the sloshing mass (see also figure~\ref{fig:design_model}). There is no definite physical cause behind $F_d$. The input, together with the output from relative velocity of slosh mass and tank wall ($\dot{\Delta}$, see equation~\ref{eq:C}), specifies a disturbance response of the closed loop system towards an artificial disturbance force $F_d$. This weighting will be used to specify a desired damping speed. The second output from the acceleration measurement can be used to put a limit on the actuation commands. It will also give a handle on the closed loop bandwidth.

Figure~\ref{fig:weightings} illustrate the relation between the two weighting functions, $W_d$ for the disturbance, and $W_c$ for the control effort. If the goal is to damp the oscillation which is at $\omega_o = 4.4 \ rad/sec $ within two cycles ($2 \cdot 1.4 \sec$) the corner frequency of the disturbance weighting should approximately be at $\omega_d = 2.2 \ rad/sec$. The roll-off $\omega_c$ will be selected about $d=20 \ rad/sec$ away. By formulating the weightings relative to $\omega_o$ the controller design can easily iterated for other filling levels and even tank shapes which at first order only differ in other $\omega_o$ (apart from $m_s$ and $m_r$ which is covert in the plant model ).

\subsection{$\mu$ Design}
Structured singular value analysis and design is a well establish method for designing robust controller (see for example~\cite{zhou:robust}). Details concerning linear fractional representation (LFT) can be found in~\cite{journals/ejcon/HeckerV04}. $\mu$ design deals with robust stability as well as with robust performance. 

\begin{figure}[!htb]
\centering\fbox{
\includegraphics[width=0.7\textwidth]{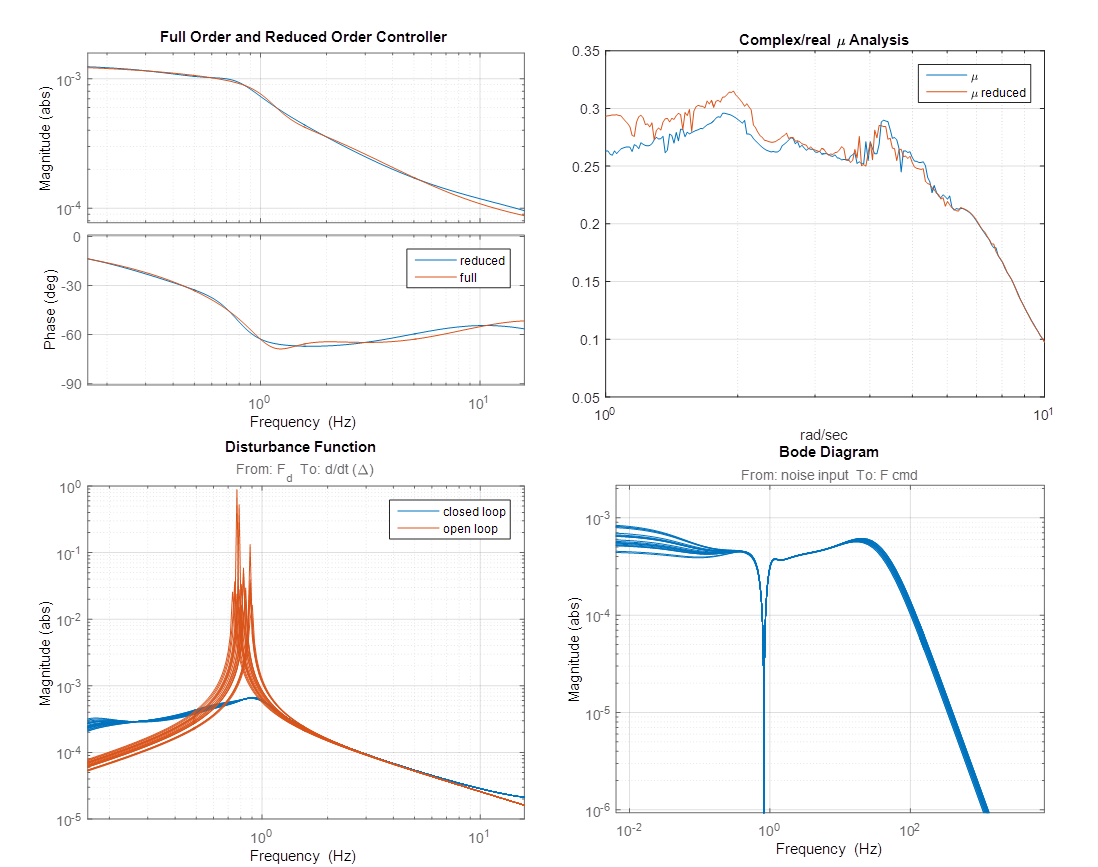}}
\caption{Top left: Full and reduced order controller transfer function (TF). Top right: $\mu$ value for full and reduced order.
Bottom left: Disturbance TF open and closed loop. Bottom right: TF measurement noise to controller command
}\label{fig:mu_ctr}
\end{figure}

The emphasis in the current activity was robust stability. The goal was to examine whether active damping is principally feasible.  The achievable performance was seen as the outcome of hardware in the loop test as opposed to be an a-priori requirement.
This is reflected in the value of the structured value which is well below one. Figure~\ref{fig:mu_ctr} gives an overview of the controller and the closed loop characteristic. The controller as represented  in the figure was the starting point of the closed loop test with CFD as well as with real $1100 \ l$ in the loop. This conservatism was justified by the fact that  stability of the tank-in-the-loop test was of paramount importance. The low value of $0.3$ indicated that (at least theoretically) there is plenty of room for increasing the requirements and still having robust stability. However as it turned out the presented controller in figure~\ref{fig:mu_ctr} showed excellent active damping performance first in the CFD case (see figure~\ref{fig:cl11}) and later in the Hexapod case.

The transfer function of the controller is shown in the top left of figure~\ref{fig:mu_ctr}. Although the available computation power of the test set-up would have allowed for high controller order, a controller reduction has been applied just out of principal. The difference in the transfer function as well as in the $\mu$ is shown in the figure. The next section will present a different approach towards low order controller via fixed order design. 

The bottom row of figure~\ref{fig:mu_ctr} shows the disturbance rejection function for the open and the closed loop. Ten randomly selected transfer functions (TF) are shown in order to give visual indication about the parameter variation. As can be seen the closed loop is well damped. In addition the the closed loop variation is much smaller than the open loop ones. The TF for the measurement to the controller command is also shown in order to provide an information about the filter roll-off characteristic.

\subsection{Validation with CFD in the Loop}

\begin{figure}[!hbt]
\centering\fbox{
\includegraphics[width=0.7\textwidth]{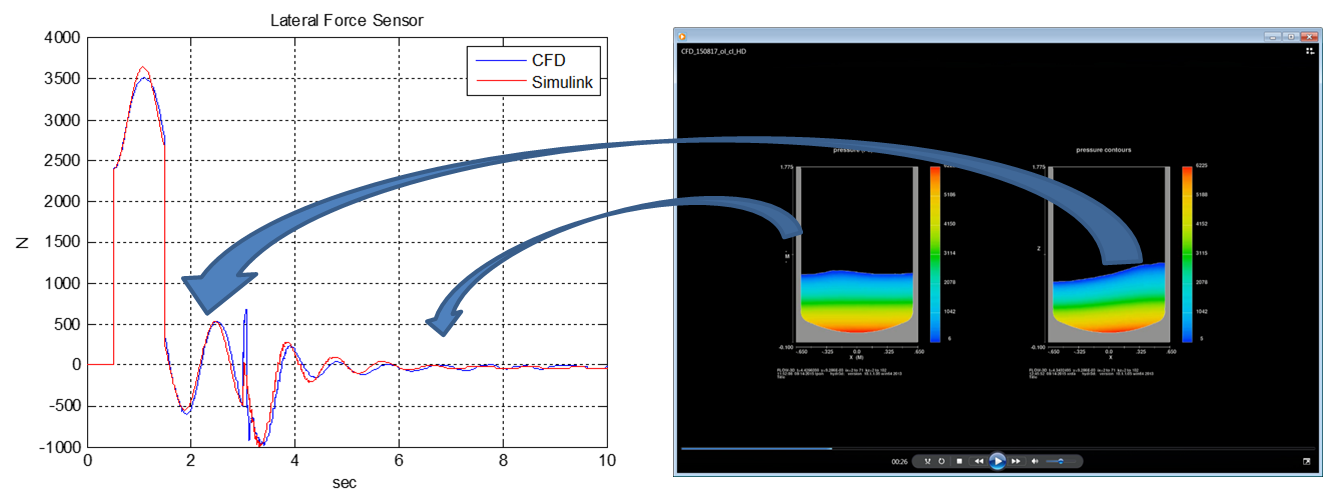}}
\caption{ $\mu$ controller with CFD in the loop (compare with figure~\ref{fig:id1})
}\label{fig:cl11}
\end{figure}

Figure~\ref{fig:cl11} shows the time domain response of the closed loop for a simulation based on the controller design plant and for the response with CFD in the loop. The figure must be compared with the open loop response of figure~\ref{fig:id1}. As can be seen the oscillation is damped within two cycles. The differences between simulation with the linear plant model, as expressed in~\ref{eq:A_mat}, and the non-linear CFD are fairly small. 

\subsection{Parametric Robust Structured Design}

The foregoing section described the control design based on $\mu$ synthesis. The validation carried out with CFD in the loop indicated the suitability of the method. Yet, it had been decided to develop an alternative controller based on a fixed structure design following~\cite{Apkarian2015}. The goal was to have an alternative at hand when it came to the hardware-in-the-loop tests and the $\mu$ based design comes out as inadequate. 

The structured singular value $\mu$ synthesis technique, as applied in the first design, has the draw back to produce a high order controller with no structure. Implementation into embedded systems often requests the need of model order reduction which can interfere arbitrarily with the optimality properties of the original $\mu$ synthesis. In addition to this it is custom in industrial practice to add non-linear elements like limiter on magnitudes, rate limiter or integrator wind-up protection to the controller. For example, in a $\mu$ synthesis based controller a possible integrator is buried into a single transfer function of the controller. The option to protect inputs or outputs of controller states by limiting them to known upper or lower bounds is also not possible in an unstructured implementation. Of course it has been a longstanding practice to tune fixed structure controller using global parameter optimization. Recently a more systematic approaches have been emerged. 

\begin{figure}[!hbt]
\centering\fbox{
\includegraphics[width=0.7\textwidth]{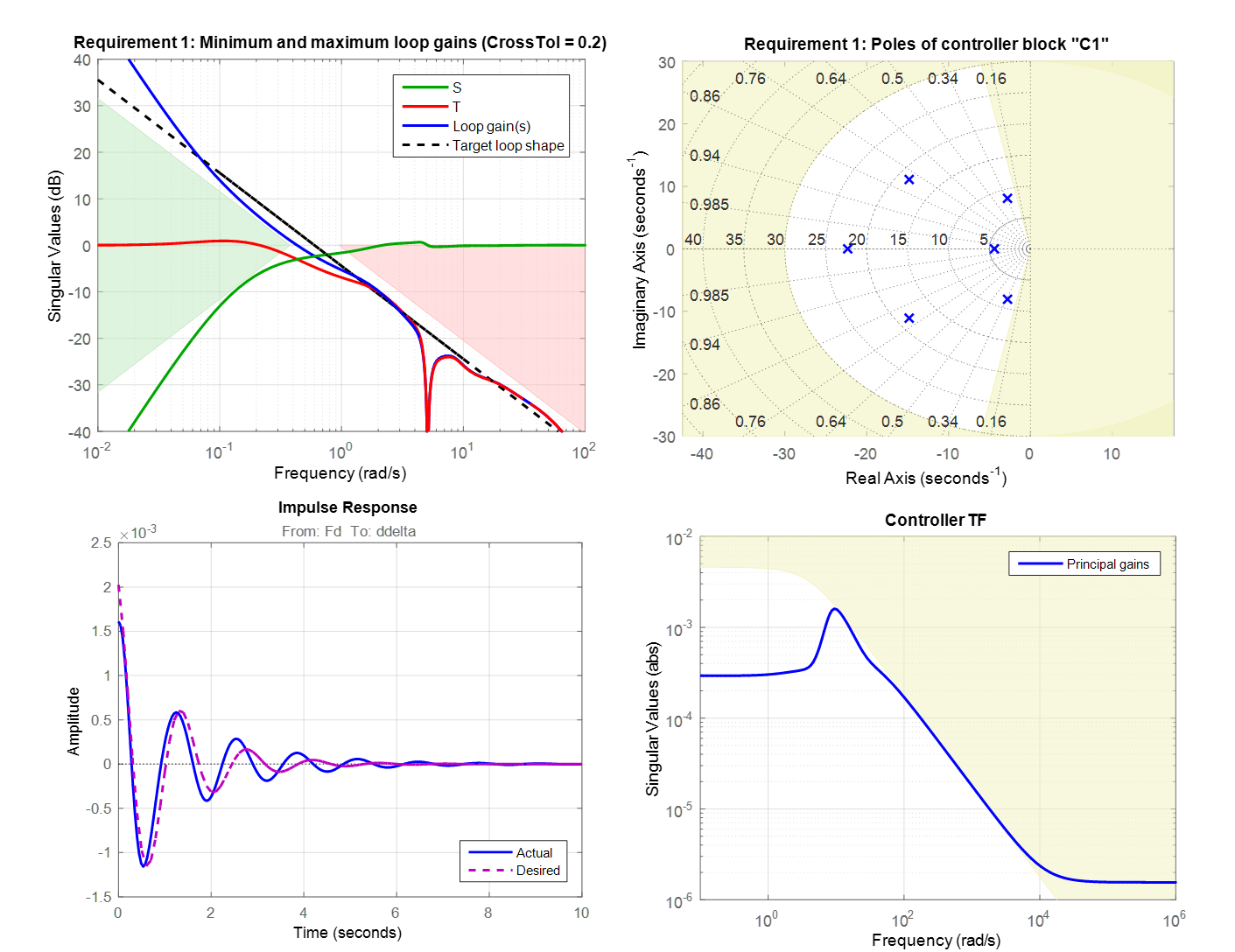}}
\caption{ Required and achieved open and closed loop specification: Top left: Open loop transfer function $L$, 
Top right: Pole location of the controller, Bottom left: Impulse response, Bottom right: controller roll-off
}\label{fig:systune}
\end{figure}

In~\cite{Falcoz2015} a fixed structure based $\mu$ synthesis involving non-smooth optimization has been applied. Basically the problem is formulated as a systematic multi-model control design problem. More recently the algorithm described in ~\citep{Apkarian2015} has become available in Mathworks' Robust Control Toolbox R2015b release. It is based on a technique to compute an inner approximation with structured controller such that a robust stability and performance is achieved for the set of uncertain parameters. This approach has been used for developing an alternative controller and is described in the following. The control structure is selected as:

\begin{equation}
C(s) = V  \frac{s^2 + 2 \zeta_1 \omega_{n} + \omega_n^2}{s^2+2 \zeta_2 \omega_{n} + \omega_n^2} 
\frac{\omega_1}{s + \omega_1} \frac{\omega_2}{s + \omega_2}
\end{equation}

The optimization search goes over the proportional gain $V$, the damping $\zeta_1$, $\zeta_2$ and the three frequencies.
The two lowpasses shall guarantee a strictly proper controller and the general notch allows to create notch or differential behavior.

\begin{figure}[!hbt]
\centering\fbox{
\includegraphics[width=0.7\textwidth]{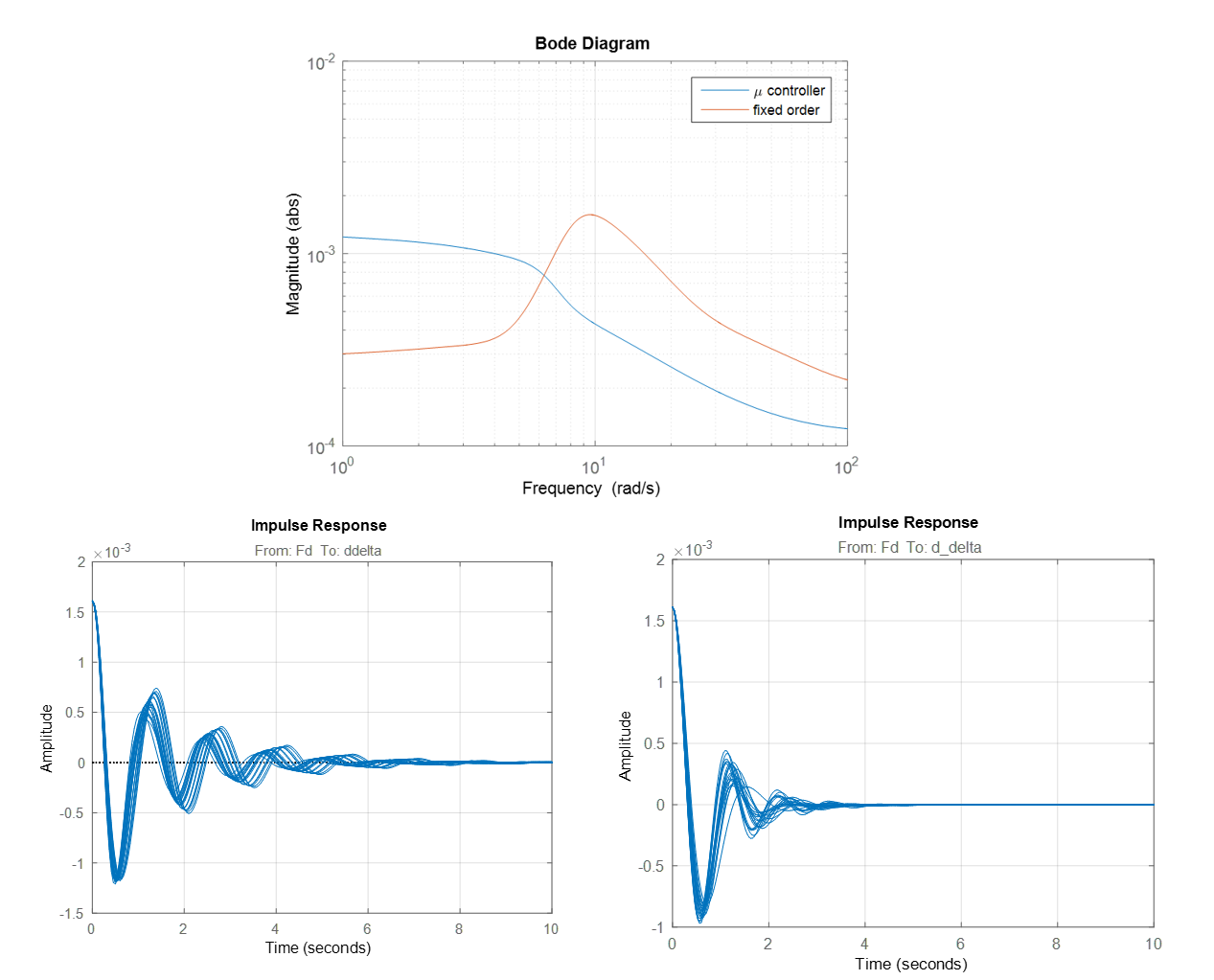}}
\caption{ Top: Controller transfer function for $\mu$ and fixed order design, Bottom left: $10$ random sample for the impulse response of the fixed order controller, Bottom right: impulse response for the $\mu$ controller.
}\label{fig:impulse}
\end{figure}

The algorithm from~\citep{Apkarian2015} allows to specify requirements in very different ways (either hard or soft). The worst case $H_\infty$ norm over a specified set of uncertain parameters (the same that have been used in the previous $\mu$ design) is then minimized over the set of parameters of the specified controller structure. Figure~\ref{fig:systune} depicts the four main requirements (not displayed is the limitation of the transfer function from disturbance to actuation (actuation effort)). The requirements cover a diverse set of possibilities. The open loop $L$ is specified in the frequency domain. That way a certain range for the  bandwidth and a proper $20db/decade$ slope at crossover (for robustness) is requested. A second specification is the allowable region for the controller poles which shall guarantee stable controllers. The third specification is the roll-off frequency for the controller which shall limit the noise amplification. Finally as a soft requirement the time domain specification is given by specifying the impulse response. The requirement is generated by computing the impulse response of the $\mu$ synthesis controller. The idea is to force the algorithm towards a closed loop behavior similar to the first design. Figure~\ref{fig:systune} shows the specified and achieved results.

Figure~\ref{fig:impulse} compares both controllers. The $\mu$ controller transfer function is basically a proportional controller with lowpass behavior starting around $8 Hz$. The fixed order controller has proportional/derivative character. The spectral energy for the $\mu$ controller is more located in the lower frequency range while the fixed order controller mostly operates around the plant bandwidth. The impulse response of both are also shown in figure~\ref{fig:impulse}. The fixed order controller has a slower response and consequently commands less actuation. This was exactly the intention for having two options at the hardware-in-the-loop test. The fixed order structure design would also allow to modify the controller quickly and transparently on the spot by frequency based tweaking of the controller parameter via the open loop $L$ transfer function. As it turned out this was not necessary at all and both controller worked very well (see figure~\ref{fig:test_2}).

\subsection{Control Architecture}
The core of the controller is simply a transfer function which represents the $\mu$ controller or the fixed structure one. However some extra layer and pre-cautions improve the overall design and are in part necessary to achieve the goal of limiting the overall travel of the Hexapod system (stroke). 

\begin{figure}[h!]
\begin{center}
\begin{tikzpicture}[>=stealth,node distance=0.5cm and 0.5cm]
\tikzstyle{sum} = [draw, circle, node distance=1.0cm and 0.5cm]
  \coordinate (orig)   	at (0,0);
  \coordinate (A0)	   	at (1,4);

  \node[draw, minimum width=1.5cm, minimum height=1.0cm, anchor=south west, text width=1.5cm, align=center] (A) at (A0) {$NL$};
  \node[sum, right=of A] (sum) {\tiny{+}};
  \node[draw, minimum width=2.0cm, minimum height=1.0cm, anchor=south west, text width=2.0cm, align=center,right=of sum,yshift=1cm] (B){upper level \\ ctr};
  \node[draw, minimum width=2.0cm, minimum height=1.0cm, anchor=south west, text width=2.0cm, align=center,right=of sum,yshift=-1cm] (C){$\mu$ ctr};
  \node[draw, minimum width=2.0cm, minimum height=1.0cm, anchor=south west, text width=2.0cm, align=center,above=of B] (D){hexapod\\ actuation};
  \node[minimum width=1.5cm, minimum height=1.0cm, anchor=south west, text width=1.5cm, align=center,right=of C,yshift = -0.2cm,xshift=1cm] (E){};  
  \node[minimum width=1.5cm, minimum height=1.0cm, anchor=south west, text width=1.5cm, align=center,left=of A] (F){}; 
  \node[minimum width=1.5cm, minimum height=1.0cm, anchor=south west, text width=1.5cm, align=center,left=of F] (G){$\ddot{x}_{cmd}$};
    
  \draw[->] (E) -- node[below right]{$F_S$}([yshift = -0.2cm]C.east);
  \draw[->] (B) -| node[above right]{$\ddot{x}_u$}(sum.north);
  \draw[->] (C) -| node[below right]{$\ddot{x}_{\mu}$}(sum.south);
  \draw[->] (sum.west) -- (A.east);
  \draw[-] (A) -- (F.center);
  \draw[->] (F.center) -- (G);
  \draw[->] (F.center) |- (D.west);
  \draw[->] (D.east) -- node [above right]{$\hat{\ddot{x}}_d$} ([xshift = 0.5cm]D.east) |- (B.east);
  \draw[->] (D.east) -- node {} ([xshift = 0.5cm]D.east) |- ([yshift = 0.2cm]C.east);
    
\end{tikzpicture}

\caption{Top Level Controller Structure: The main module is named '$\mu$ ctr'. The 'upper level ctr' is a low bandwidth control taking care that the Hexapod stroke stays in its limits. The 'hexapod actuation' contains the model of the Hexapod actuation chain. The output
is the estimation of the slosh motion $\hat{\ddot{x}}_d$. The 'NL' block contains non linearities like rate and magnitude limiters. }
\label{fig:ctr1}
\end{center}
\end{figure}
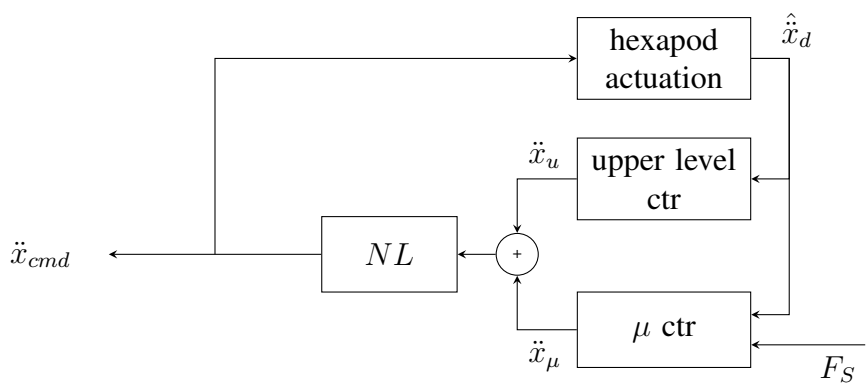

As already shown the force measurement output is triggered not only by reaction force of the sloshing propellant but also by rigid body acceleration, which originates in the controller command itself (an effect, which basically is reflected in the non-zero $D$ matrix). In order to cover this effect a kind of feed-forward switching logic is applied: The commanded output from the $\mu$ module is sent through a model of the actuation chain (here a lowpass filter and a delay). This output is subtracted from the measurement. That way an estimate of the sloshing reaction force is achieved. Likely differences between model and actual Hexapod drive system are covered with an appropriately large $\Delta$ in the LFT.

The further complication comes from the fact that the available stroke of the Hexapod is limited (approximately $\pm 0.04 \ m$). The core controller design aims at reducing the relative velocity between slosh mass and tank wall to zero (weighting $W_d$) and therefore damping the relative motion. However the inertial velocity of the of the tank itself is not taken into consideration in the $\mu$ design. Even if the $\mu$ controller can achieve the damping with the available stroke within, e.g. $2 sec$, a certain amount of terminal velocity maybe left in the system. The Hexapod would be moving with constant velocity to one side (without exciting the fluid because it has the same velocity). This phenomenon is covered by an outer loop. The outer loop is a velocity feedback of the the tank itself with a much lower bandwidth than the inner $\mu$ controller. 

In fact there is even a further precaution with the aim to keep the damping action within the $0.04 \ m$. Although the damping works regardless of the phasing of the start of the controller with regard to the oscillatory motion, the stroke is minimized in case the controller begins when the reaction force is at the maximum. This timing logic is not shown in figure~\ref{fig:ctr1} in order to avoid cluttering the schematic.

\section{Physical Modeling and Embedded Software Generation}

\begin{figure}[!ht]
\centering\fbox{
\includegraphics[width=0.7\textwidth]{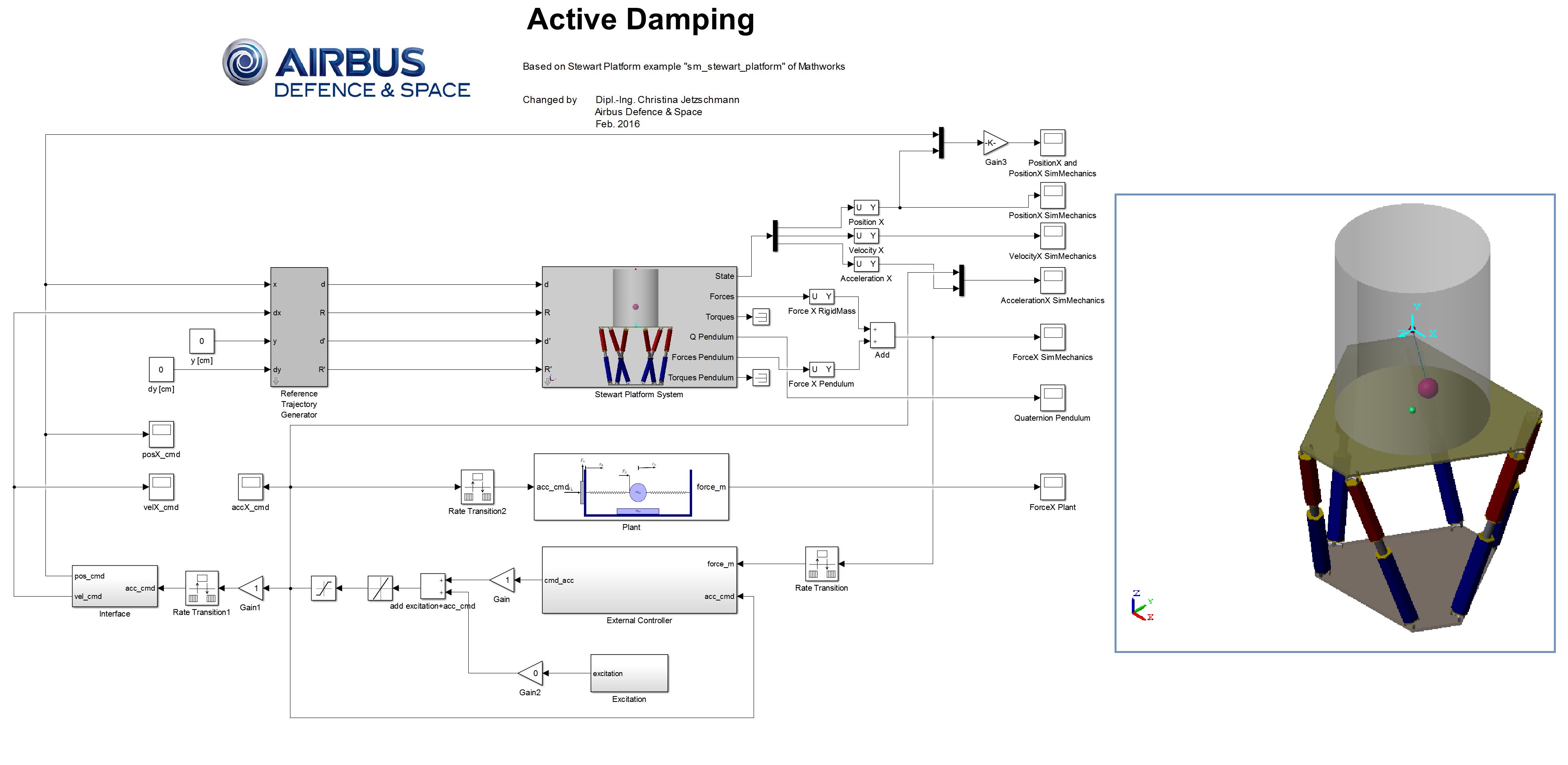}}
\caption{Simulator with plant dynamics realized either as SimMechanics based physical modeling 
(top block) and Simulink based differential equation (DEQ) (middle block). System can run in parallel. Right hand side: SimMechanics Hexapod visualization.
}\label{fig:SM_ActiveDamping}
\end{figure}

A unified simulation environment is based on Simulink. Two physical models are embedded: The Hexapod/pendulum via SimMechanics and the fluid motion via a CFD simulation (Flow3D) (see figure~\ref{fig:SM_ActiveDamping}). The controller (green block) can drive a plant model realized as differential equation, a pendulum attached to a Hexapod, as well as a coupled system which consist of Flow3D for the fluid part and the (then empty) tank on the Hexapod. 
Results reported in Figure~\ref{fig:cl11} were created with this environment. 

\begin{figure}[!hb]
\centering\fbox{
\includegraphics[width=0.7\textwidth]{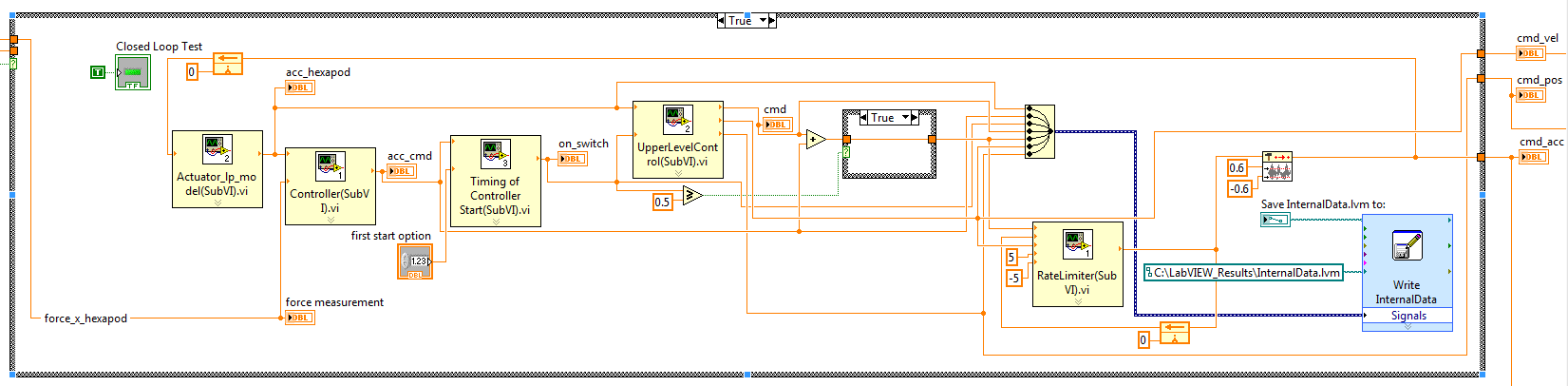}}
\caption{Labview Representation of the controller.
}\label{fig:ClosedLoopTest}
\end{figure}

This unified simulation allowed an iterative build-up of the simulation starting with the DEQ and a perfect actuation system and ending with the hexpod drive system acting on a CFD representation. In principle it would be possible to use auto-code for the controller to further unify the development chain into the embedded system of the real hardware. This last step was not possible to implement, because the existing Hexapod system used a Labview based system (see figure~\ref{fig:ClosedLoopTest}). However, for the dedicated control architecture script files where developed which allowed an automated transfer of the control parameter directly from the Matlab design files to the LabView system. This allowed quick and reliable updates of the controller during tests. 

\section{Hexapod Test}
\begin{figure}[!hbt]
\centering\fbox{
\includegraphics[width=0.7\textwidth]{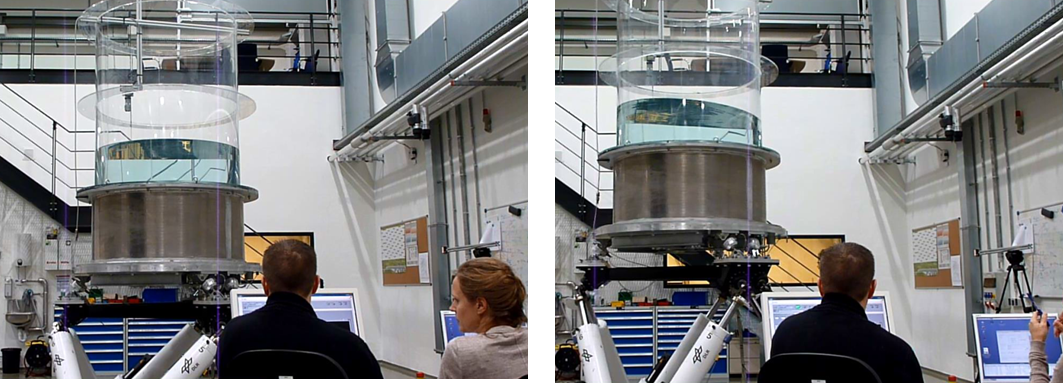}}
\caption{ Hexapod Operation at the Cryo-Lab of the Institute of Space Systems of the German Aerospace Center (DLR) in Bremen (~\cite{Gerstmann2015})
}\label{fig:Hexapod_fig}
\end{figure}

Figure~\ref{fig:Hexapod_fig} shows the Hexapod system. The Hexapod from Bosch Rexroth consists of six actuators. Each actuator contains a piston, pneumatic driven with a spindle. On top of the actuators is a steel frame to flange different kinds of setups. For the experiments, an experimental platform  with six force transducers is installed. 
The experimental platform can be tilted up to $20 \ deg $. The payload can be accelerated up to $0.6 \ g$. The forced motion can be performed with a maximum excitation amplitude of $200 \ mm$ and frequency of $10 \ Hz$. To measure the status of the liquid inside the tank a pair of six HBM U10M-5kN transducers is used (for more information see~\citep{Konopka2016}).

\begin{figure}[!hbt]
\centering\fbox{
\includegraphics[width=0.7\textwidth]{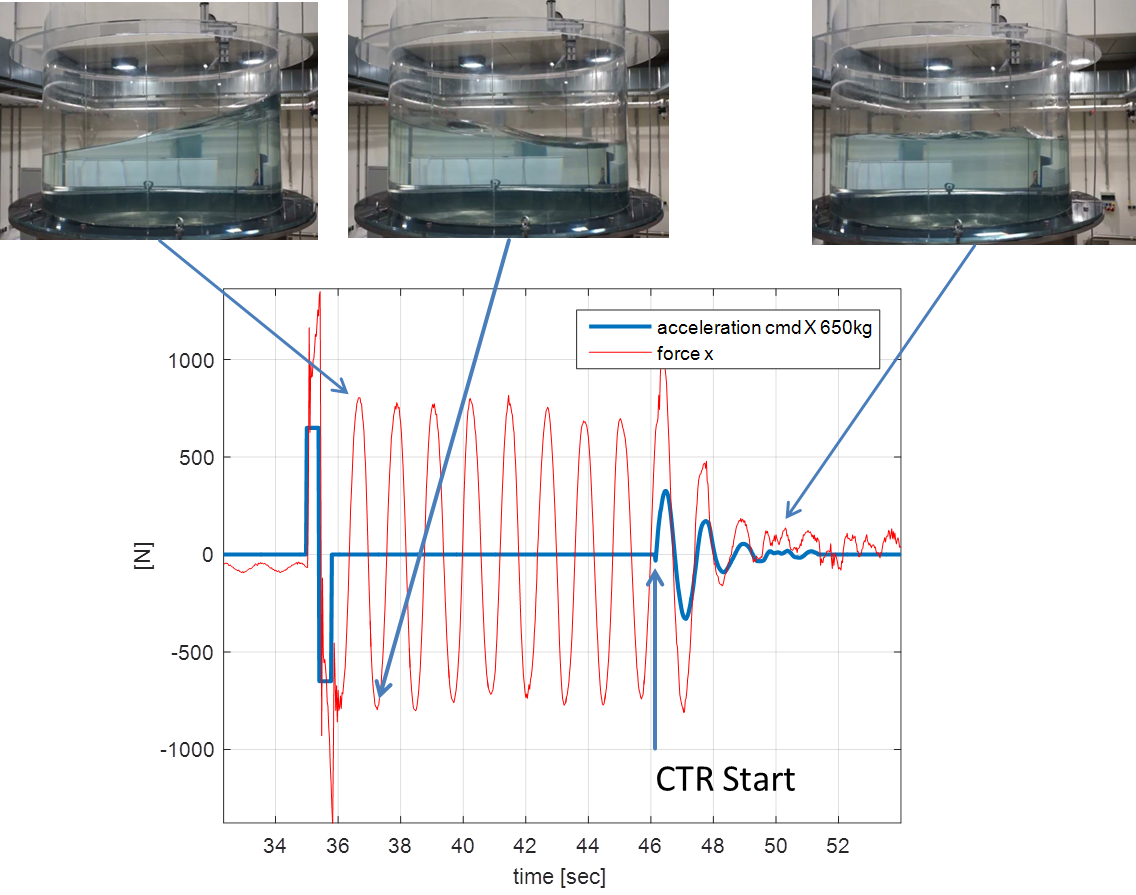}}
\caption{ Damping Example for $600$ liter and pulse excitation.
}\label{fig:slash}
\end{figure}

Tests were performed with $600$ and $1100 $ liter with different level and types of excitation. The $600 \ liter$ case was carefully prepared in coupled simulation with CDF (see figure~\ref{fig:cl11}) based on system identification of open loop experiment (see section~\ref{sysid}). For the $1100$ liter case the interpolation formulas from~\cite{dodge2000} were applied.

Figure~\ref{fig:slash} illustrates one test with a step-wise acceleration. A short pulse was applied in order to excite the fluid (at $t=35 \ sec$). The water was left in free motion for about $10  sec$. As can be seen from the movie snapshot the fluid reached an angle of more the $45 \ deg$. Second order harmonics can also be seen. At $t=46 \ sec$ the controller starts and the motion is nearly completely damped after two cycles. Comparing this with the CFD simulation as shown in figure~\ref{fig:ClosedLoopTest} one can observe a slightly slower damping on the Hexapod. This can be attributed either to un-modeled delays or imperfections in the Hexapod driving mechanism (albeit the $\Delta$ LFT frameworks captures these effect as uncertainties) or due to non-linearities in the real fluid motion. In any case the closed loop performance is satisfactory.

\begin{figure}[!hbt]
\centering\fbox{
\includegraphics[width=0.7\textwidth]{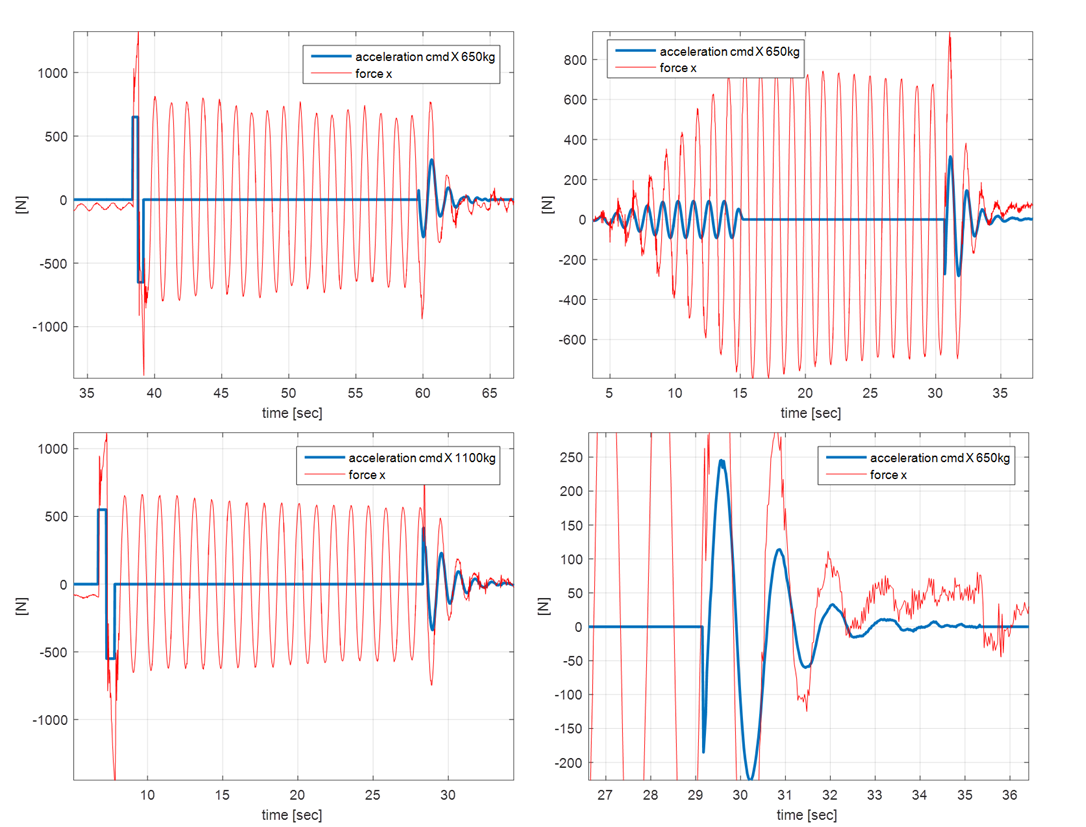}}
\caption{ Top left: $600$ liter, pulse type excitation, Top right: $600$ liter case, sine excitation,
Bottom left: $1100$ liter case, Bottom right: Zoom into the controller phase
}\label{fig:test_1}
\end{figure}

Figure~\ref{fig:test_1} displays results for a different excitation. The fluid is exposed for about $10 \ sec$
to a forced motion close to the eigenfrequency (top left figure). The top right figure shows a case where the free motion was longer than in figure~\ref{fig:slash} in order to analyze the influence of a different initial condition at controller start. The bottom left figure displays a result for the $1100$ liter case.

\begin{figure}[!hbt]
\centering\fbox{
\includegraphics[width=0.8\textwidth]{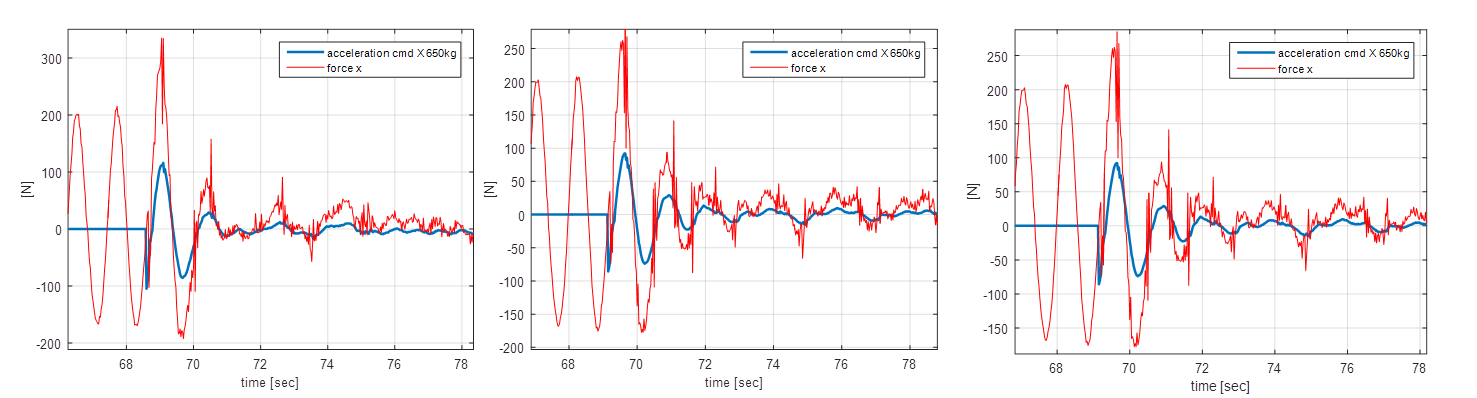}}
\caption{ Different controller: Left - $\mu$ controller, Middle - Fixed order, Right - Fixed order with higher uncertainty
}\label{fig:test_2}
\end{figure}

Figure~\ref{fig:test_2} illustrates the differences when the controller, derived from $\mu$ design, and the controller, based on parametric structured design, is used. As already indicated by the comparison of the impulse response of the two controller (figure~\ref{fig:impulse})
the fixed structure controller takes one more cycle for damping. Both controller achieve the goal of quickly damping the slosh motion.

\section{Conclusions}
The paper described the whole development and testing cycle for system which actively damps the sloshing motion. Based on the measurement of the reaction force of the fluid the container is accelerated such that the sloshing motion is damped within a time frame which corresponds to about two oscillatory cycles. The presented activity proved that, following a model based design approach, it is possible to actively influence fluid motion. The applied controller design techniques led to a robust closed loop behavior. The closed loop behavior with $1100 liter$ in the loop was very similar to the predicted one with the CFD in the loop. Even the design model provided a good prediction. 

A classical structured singular value synthesis controller and fixed structure design based on a recently developed algorithm from~\citep{Apkarian2015} has been used. Both controller were able to achieve a very good damping.

\clearpage

\bibliographystyle{ieeetran}

\bibliography{refs}

\end{document}